\documentclass[%
 reprint,
groupedaddress,
nofootinbib,
 amsmath,amssymb,
 aps,
 prstab,
floatfix,
]{revtex4-2}
\usepackage{color}
\usepackage{graphicx}
\usepackage{dcolumn}
\usepackage{bm}
\usepackage{siunitx}
\usepackage{multirow}
\usepackage{physics}
\usepackage{bookmark}
\usepackage{comment}

\begin{document}


\title{Synchrotron radiation leveling at future circular hadron colliders}

\author{F. Zimmermann} 
\email{frank.zimmermann@cern.ch}
\affiliation{CERN, 1211 Geneva 23, Switzerland}



\date{\today}

\begin{abstract}
Luminosity leveling to limit the event pile up
is a key ingredient of the LHC luminosity upgrade,
the High-Luminosity LHC (HL-LHC).
For a future circular hadron collider, such as the FCC-hh, 
operating at a centre-of-mass energy
of 70--90 TeV, synchrotron radiation 
becomes significant, with radiation 
damping times of the order of one or a few hours.
The rapid shrinkage of the emittance may call for a
leveling of the beam-beam tune shift or of the
event pile up, 
as previously explored.
However, the strong synchrotron radiation emitted inside 
the cold superconducting magnets also 
represents a significant heat load and 
is likely to limit the total beam current. 
In this article, we discuss a new approach, namely 
synchrotron radiation power leveling, where the 
beam energy is adjusted during a physics store,
either continually or in a few discrete steps, while the beam current decreases, 
so as to keep the synchrotron radiation power at or below a certain limiting value. 
In this way, both peak and integrated luminosity of the FCC-hh are increased,
compared with operation at a fixed beam energy.  
The FCC-hh detectors, and in particular the physics event analysis, 
need to be prepared for this novel mode of operation.
This article presents two example running scenarios 
for synchrotron radiation leveling at the FCC-hh. 
While not greatly reducing the integrated luminosity at highest
collision energy, synchrotron-radiation leveling 
can significantly increase the number of events for key processes already occurring 
at lower energy. As an example,  we show that it
raises the number of di-Higgs production events 
by 60\% or more.

%

\end{abstract}

\maketitle

\section{\label{sec:intro}Introduction}
At the presently operating Large Hadron Collider (LHC) \cite{lhc} and, especially, its High-Luminosity upgrade, the HL-LHC \cite{hllhc}, the luminosity during a physics run 
is, or will be, leveled 
--- through changing crossing angle, interaction-point beta function, beam separation, 
or crab-cavity voltage --- in order to limit the 
event pile up in the detectors \cite{lev1,lev2,lev3,PhysRevSTAB.17.111001,Gorzawski:2207366,8673806,Hostettler:2023uav}. 
At these existing or soon-to-be-commissioned hadron machines, 
the effect of synchrotron radiation on the transverse emittance during a physics store is still rather negligible and easily overshadowed by emittance growth from intrabeam scattering \cite{Papadopoulou:2813519}, 
power-converter ripple \cite{Kostoglou:2841809}, 
crab-cavity noise \cite{PhysRevAccelBeams.27.051001}, etc.  

For a future ``100-TeV''-scale collider, such as the proposed Future Circular hadron Collider, FCC-hh \cite{fcchh,fcchh1,Benedikt:2952712,PerezSegurana:2926920},  
the transverse emittance 
damping time is likely to be  
shorter than the proton  burn-off time. 
As a result, at such a machine, the total beam-beam tune shift
increases during the store. 
In Ref.~\cite{PhysRevSTAB.18.101002}, we derived analytical expressions for the optimized integrated luminosities and for 
the optimum store lengths at high-energy hadron colliders limited
either by the maximum acceptable event pile up or by an empirical 
beam-beam tune shift limit, 
in the presence or absence of strong radiation damping.
Our analytical solutions were 
illustrated with examples from the existing LHC, for the 
planned HL-LHC and for the proposed FCC-hh.  
By contrast, at the lower-energy HE-LHC \cite{Zimmermann:2651305}
the proton burn off time would be slightly shorter than 
the radiation damping time. The optimum run time and
optimum luminosity for this situation
were presented in Ref.~\cite{BENEDIKT2018200}.

For the FCC-hh, the strong synchrotron radiation emitted inside 
the cold superconducting magnets, shielded by a beam screen,
represents a significant heat load, and either through 
the total power consumption or by the limited 
capacity of the installed cryogenic systems, 
it is likely to limit the total beam current. 

For this situation, synchrotron radiation power leveling becomes a natural choice to maximize the physics output \cite{lipeles}.
Here, the beam energy is adjusted during a physics store, either continually or in a few discrete 
steps, while the beam current decreases due to proton burn-off in collision, 
so as to keep the synchrotron radiation power at or below a certain limiting value. 
Synchrotron radiation leveling allows for much higher initial beam currents than when starting with collisions at the maximum beam energy.    
In this way, both peak and integrated luminosity of the FCC-hh are increased,
compared with operation at one fixed beam energy. 

The key parameters for FCC-hh are compiled in Table~\ref{hhfuturetable}, considering 
a fixed arc dipole field of either 14 or 12 T 
(cases ``FCC-hh-14'' and ``FCC-hh-12''), in preparation for our later discussion.  
The table also illustrates how the synchrotron
radiation strongly increases 
for the higher magnetic fields.  

\begin{table*}[htbp]
\caption{Parameters of FCC-hh at 12 T and 14 T dipole field,  
compared with the HL-LHC and LHC. For the integrated luminosity, 160 days of operation per year, and 
70\% machine availability is assumed~\protect\cite{Bordry:2645151}. 
All three colliders have a regular 
bunch spacing of 25 ns.  
\label{hhfuturetable}}
\centering 
\begin{tabular}{lccccc}
\hline\hline
 & Unit & FCC-hh-12 & FCC-hh-14 & HL-LHC & LHC \\
\hline
Centre-of-mass energy & TeV &
72 & 84 & \multicolumn{2}{c}{14} \\ 
Circumference & km & \multicolumn{2}{c}{90.7} & \multicolumn{2}{c}{26.7} \\
Dipole field & T & 12 & 14 & \multicolumn{2}{c}{8.33} \\
Beam current & A & 0.93 & 0.5 &  1.1 & 0.58 \\
Bunch Intensity & $10^{11}$~p
& 1.85 & 1.0 &  2.2 & 1.15 \\
No.~bunches / beam  &--- & 9500 & 9500 & 2760 & 2808 \\
Total synchr.~radiation power & kW & 2400 & 2400 & 15 & 7  \\
SR power / length & W/m/aperture & 
15.6 & 15.6 & 0.33 & 0.17 \\
Longit.~emit.~damping time $\tau$/2 & h  & 1.2 & 
0.75 & \multicolumn{2}{c}{12.9} \\
IP beta function $\beta_{x,y}^{\ast}$ & m & 0.3 & 0.3 &  0.15 (min.) & 0.55 \\
Initial normalised rms emittance & \si{\micro \meter} &
2.5 & 2.2 &  2.5 & 3.75 \\
No.~high-lum.~interaction points $n _{\rm IP}$ & --- &
 2  & 2 & 2  & 2 \\
Initial total b.-b. tune shift $\Delta Q _{\rm tot}$ &  &
0.018 & 0.011 &  0.015 &  0.01 \\
Max.~total b.-b. tune shift $\Delta Q _{\rm tot,max}$ &---  &
0.025 & 0.025 &  0.015 &  0.01 \\
Initial luminosity & $10^{34}$~cm$^{-2}$s$^{-1}$ & 
44 & 16  &  5 (lev.) & 1 \\
Peak luminosity & $10^{34}$~cm$^{-2}$s$^{-1}$ & 
 44 & 20  &  5 (lev.) & 1 \\
Inelastic cross section $\sigma_{\rm inel}$ & mb & 105 
  &  107  &  83 &  83 \\
Total cross section $\sigma_{\rm tot}$ & mb & 148 
  & 151  & 112   &  112 \\
Initial no.~events / bunch crossing $\mu$ & --- & 
1460  & 580 
 & 132 & 27 \\
Peak no.~events / bunch crossing $\mu$ & --- & 2050 & 1300 
 & 132 & 27 \\
Initial proton burn up time $\tau_{\rm bu}$  & hr & 4.8 
  & 5.2  &   &   \\
Stored energy / beam & GJ & 10.3  & 6.5
& 0.7 & 0.36 \\ 
Average turnaround time  $t_a$ & h & 5  & 5
& 4 & 5 \\ 
Accelerator availability & --- & 70\%  & 70\% 
& 80\%  & 78\% \\ 
Int.~annual luminosity / interaction point & ab$^{-1}$/y & 1.3 & 0.7  & $\ge$0.25 & 0.05  
\\
\hline\hline
\end{tabular}
\end{table*}

In the next section, we first review and simplify the main equations describing the 
evolution of bunch intensity, emittance, beam-beam tune shift,
and luminosity during a physics store, assuming a maximum acceptable or achievable beam-beam tune shift, and then evaluate the resulting optimum integrated luminosity.   
We provide numerical examples for the cases FCC-hh-12 and FCC-hh-14 from Table \ref{hhfuturetable}.    

In the following, we present two scenarios for physics operation with synchrotron radiation 
leveling, at or between these two energies. 
The first considers running, in each physics store, at two different beam energies, namely 
one lower and the other equal to the maximum design energy.
The second scenarios presents a case of continuous energy change during each store.
We compare the performance with those obtained when running 
only at a fixed energy, corresponding to an arc dipole field of either 12 T or 14 T. 

\section{\label{sec:bd}Constant energy dynamics}
In this section, considering the conventional case of operating at one constant beam energy,
we recall and slightly 
simplify the relevant formulae of Ref.~\cite{PhysRevSTAB.18.101002}.  

We consider the collision of round Gaussian beams with equal horizontal and vertical geometric rms 
emittances, $\varepsilon\equiv \varepsilon_x=\varepsilon_y$, 
and equal beta functions at the collision point, 
$\beta^{\ast}\equiv \beta_x^{\ast}=\beta_{y}^{\ast}$.
Further we assume the presence of crab cavities around each collision point 
(i.e.~no luminosity loss from a finite crossing angle), 
and we also neglect the hourglass effect (which is permitted for 
$\beta^{\ast}$ values larger than the rms bunch length).
Under these assumptions --- which, for example, 
are well fulfilled for the parameters of the FCC-hh (and also HL-LHC) ---  
the luminosity $L$ at a collision point is given by 
\begin{equation}
L = \frac{f_{\rm rev} n_{b} N_{b}^{2}}{4 \pi \beta^{\ast}\varepsilon }\; . 
\label{lumi} 
\end{equation}
where $f_{\rm rev}$ denotes the revolution frequency, $N_{b}$ the bunch population (the number of 
protons per bunch), $n_{b}$ the number of bunches per beam, and 
$\varepsilon$ the geometric rms emittance.
The initial luminosity at time $0$ is denoted by $L_0$. 
 
We also suppose that there are $n_{\rm IP}$ high-luminosity 
collision points around the ring
circumference, and, for simplicity, 
that the luminosity is equal for all of them.

Under the same assumptions as made for the luminosity, the total beam-beam 
parameter (or the approximate beam-beam tune shift) becomes 
\begin{equation}
\Delta Q_{\rm tot} = n_{\rm IP}  \frac{r_p N_b}{4\pi \gamma \varepsilon} \; ,
\label{eq:dQtot} 
\end{equation}
where $r_{p}$ designates the ``classical proton radius'' 
(about $1.5\times 10^{-18}$~m), and $\gamma$ the relativistic Lorentz factor.
Here we distinguish between the initial value of the total beam-beam tune shift $\Delta Q_0 \equiv  
\Delta Q_{\rm tot} (0)$, and the maximum acceptable value $\Delta Q_{\rm max}$.

The number of inelastic scattering events per bunch collision $\mu$ (approximately equal to 
the event pile up in the particle physics detectors) can be calculated
from the formula
\begin{equation}
\mu \equiv \sigma_{\rm inel} \frac{L}{n_{b} f_{\rm rev}} 
= \sigma_{\rm inel} \frac{N_{b}^{2}}{4 \pi \beta^{\ast}\varepsilon} 
\; ,
\label{pileup} 
\end{equation}
where $\sigma_{\rm inel}$ refers to the (inelastic) 
cross section for any event ``seen'' by the particle-physics experiment.  

If we take the dominant source of beam loss to be due 
to the burn-off in collision
(a good assumption for machines like the LHC and HL-LHC), 
the rate of change of the bunch intensity is proportional to the instantaneous 
luminosity as   
\begin{equation} 
\frac{dN_b}{dt}=-\sigma_{\rm tot} n_{\rm IP}  \frac{L}{n_b} 
= - K \frac{N_b^2}{\varepsilon}\; , 
\label{dnbdt} 
\end{equation}
where $\sigma_{\rm tot}$ denotes the total 
cross section (comprising inelastic and elastic components), 
and $L$ the luminosity at each IP (assumed to be the same), 
and, for later use, we have introduced the parameter $K$:
\begin{equation}
K \equiv \frac{\sigma_{\rm tot} f_{\rm rev} n_{\rm IP}}{4\pi \beta^{\ast} } \; . 
\label{kdef}
\end{equation}
The character of the solution of Eq. 
(\ref{dnbdt})
differs according to the constraints and assumptions imposed.
In particular it depends on the behavior of the emittance.

The energy-dependent total and inelastic cross sections, 
$\sigma_{\rm tot}$ and $\sigma_{inel}$, 
can be estimated from  scaling laws 
\cite{dominguez,cross1,cross2,silva}.  
Cross-section values for different energies
are shown in Table \ref{hhfuturetable}.

Damping due to synchrotron radiation is significant at the FCC-hh, both transversely and even more longitudinally. In the following we 
assume that the bunch length and the longitudinal emittance 
are kept constant via continuous controlled excitation with ``pink noise,''
which counteracts the natural emitttance shrinkage  
\cite{heac,dominguez}.
Such automated continuous   
longitudinal excitation is routinely applied during the energy ramp  
of the LHC \cite{pb}.

To a good approximation the initial 
transverse emittance evolution is described by an exponential decay due to radiation 
damping:   
\begin{equation}
\varepsilon (t)  \simeq \varepsilon_0 \exp \left( -\frac{t}{\tau} \right)\; .  
\label{emit1}
\end{equation} 
Inserting (\ref{emit1}) into (\ref{dnbdt}), 
the time-dependent bunch intensity fulfils 
\begin{equation}
\frac{dN_b}{dt} = - K \frac{N_b^2}{\varepsilon_0} e^{t/\tau} \; .  
\label{dnbdt2} 
\end{equation}
The integration of (\ref{dnbdt2}) yields  
\begin{equation}
N_b (t) = \frac{N_0}{B  \left( 
\varepsilon^{t/\tau}-1 \right) + 1}\; ,
\label{nevol} 
\end{equation}
where we have abbreviated the initial bunch intensity as $N_{0}\equiv N_{b}(0)$ and introduced 
\begin{equation}
B \equiv \frac{K \tau N_0}{\varepsilon_0}\; .
\label{bdef}
\end{equation}

The emittance continues to shrink until a maximum acceptable value for the total
beam-beam tune shift $\Delta Q_{\rm max}$ (the ``beam-beam limit'') is reached at a time $t_1$. 
By definition, this time $t_{1}$ follows from the tune shift limit as
\begin{eqnarray}
\Delta Q_{\rm max} & \stackrel{!}{=} & 
 \Delta Q (t_1 ) = 
n_{\rm IP}  \frac{r_p N_b (t_1)}{4\pi \gamma \varepsilon (t_1)} 
\nonumber \\ 
& = & \Delta Q_0 \, \frac{e^{t_{1}/\tau}}{B  \left( e^{t_{1}/\tau}-1 \right) + 1}\; ,  
\label{dqlim} 
\end{eqnarray}
where
\begin{equation}
\Delta Q _0 \equiv n_{\rm IP}  \frac{r_p N_0}{4\pi \gamma \varepsilon_{0} }  \; .
\label{adef}
\end{equation}

From (\ref{dqlim}) we deduce the time limit for period 1:
\begin{equation}
t_{1} = -\tau  \ln 
\left[ \frac{ 
\frac{\Delta Q_0}{\Delta Q_{\rm max}}-B
}{1 - B}
\right] \; .
\label{t1}
\end{equation}

The integrated luminosity for period 1 reads  
\begin{eqnarray}
\Sigma_1 & = & L_0 \tau \left( \frac{\Delta Q _{\rm max}/ \Delta Q_0 -1}{1-B } \right)   \; . 
\label{i1int}
\end{eqnarray}

From time $t{_1}$ onward, the collisions continue with constant tune shift. 
\begin{equation}
\Delta Q_{\rm max}=n_{\rm IP}  \frac{r_p N_b (t)}{4\pi \gamma \varepsilon (t)}\; \; \;  
{\rm for} \; \;   t>t_1\; .
\label{dqc}
\end{equation}  
This can be achieved by controlling the emittance $\varepsilon (t)$ 
through transverse noise excitation.

From (\ref{dnbdt}) the further change in intensity is
\begin{equation}
\frac{dN_b}{dt} = 
-\sigma_{\rm tot} n_{\rm IP}  \frac{L}{n_b} 
=-\frac{N_b}{\tau_2} \; , 
\end{equation}
where we have introduced the time constant
\begin{equation} 
\tau_2 \equiv \frac{\beta^{\ast} r_p}{\sigma_{\rm tot} f_{\rm rev} 
\gamma \Delta Q_{\rm max}} \; ,
\label{tau2}
\end{equation}

The solution is an exponential decay for both the bunch intensity and the emittance, that is
\begin{equation}
\varepsilon (t) = \varepsilon_0 e^{-t_1/\tau} e^{-(t-t_1)/\tau_2 }
\end{equation} 
and
\begin{equation}
N(t) =\frac{N_0}{B ( e^{t_1/\tau}-1 )
+ 1} e^{-(t-t_1 )/\tau_2 }
\label{n2evol} 
\end{equation}

The time-dependent luminosity then becomes 
\begin{equation}
L(t) = \frac{n_b }{\sigma_{\rm tot} \tau_2 n_{\rm IP}}  
N(t)
\end{equation}
or
\begin{equation}
L(t)= L_{\rm max}  e^{-(t-t_1 )/\tau_2 }
\label{ltinst}
\end{equation}
with
\begin{equation}
L_{\rm max}\equiv \frac{n_b N_0 }{\sigma_{\rm tot} \tau_2 n_{\rm IP}} 
\frac{1}{B ( e^{t_1/\tau}-1 )+ 1} 
\label{lmax}
\end{equation}

The integrated luminosity at time $t$ ($t>t_{1}$) is 
\begin{eqnarray}
\label{i1i2}
\Sigma(t) & =& \Sigma_1+\Sigma_2 (t) = \Sigma_1+\int_{t_1}^t L(t)  dt  \\ 
& = & 
L_0 \tau \left( \frac{\frac{\Delta Q _{\rm max}}{\Delta Q_0} -1}{1-B } \right) 
+ L_{\rm max} \tau_2 \left( 1 - e^{-\frac{t_r -t_1 }{\tau_2}} \right) \nonumber
 \; . 
\end{eqnarray}

Finally,  we want to maximize the average luminosity. 
Introducing  the ``turnaround time'' $t_{\rm ta}$, i.e.~the average time between the end of one 
physics fill at time $t_{r}$ 
and the start of the luminosity production in the following fill, at (on average) the
time $t_{r}+t_{\rm ta}$, 
the average luminosity depends on the run time $t_r$ 
as
\begin{equation}
L_{\rm ave} (t_r )=\frac{\Sigma_1+\Sigma_2 (t_r )}{t_r+t_{\rm ta}  }\; .
\label{lavetr}
\end{equation}

The optimum run time $t_{r,{\rm opt}}$  is found by maximizing $L_{\rm ave}$.

We define the auxiliary parameters
\begin{eqnarray}
C & \equiv & 
L_0 \tau \left( \frac{\frac{\Delta Q _{\rm max}}{\Delta Q_0} -1}{1-B } \right) 
+ L_{\rm max} \tau_2 
\nonumber
\\
D & \equiv & - L_{\rm max} \tau_2 e^{t_1/\tau_2} \; .   
\label{Ddef}
\end{eqnarray}

From 
\begin{equation}
\frac{d}{dt_r } \left( 
\frac{C+D e^{-t_r/\tau_2 }}{t_r+t_a }\right) _{t_r=t_{r,{\rm opt}}}
 =0
\end{equation} 
the optimum run time follows as 
\begin{equation}
D \left( t_{r,{\rm opt}}+t_a +\tau_2 \right) 
e^{-t_{r,{\rm opt}}/\tau_2 } + \tau_2 C \stackrel{!}{=} 0 \; ,
\label{ort1} 
\end{equation}
which can be solved numerically. 

\section{\label{sec:two}Performance at fixed energy}
\begin{figure}[htb]
    \centering
    \includegraphics[width=\linewidth]{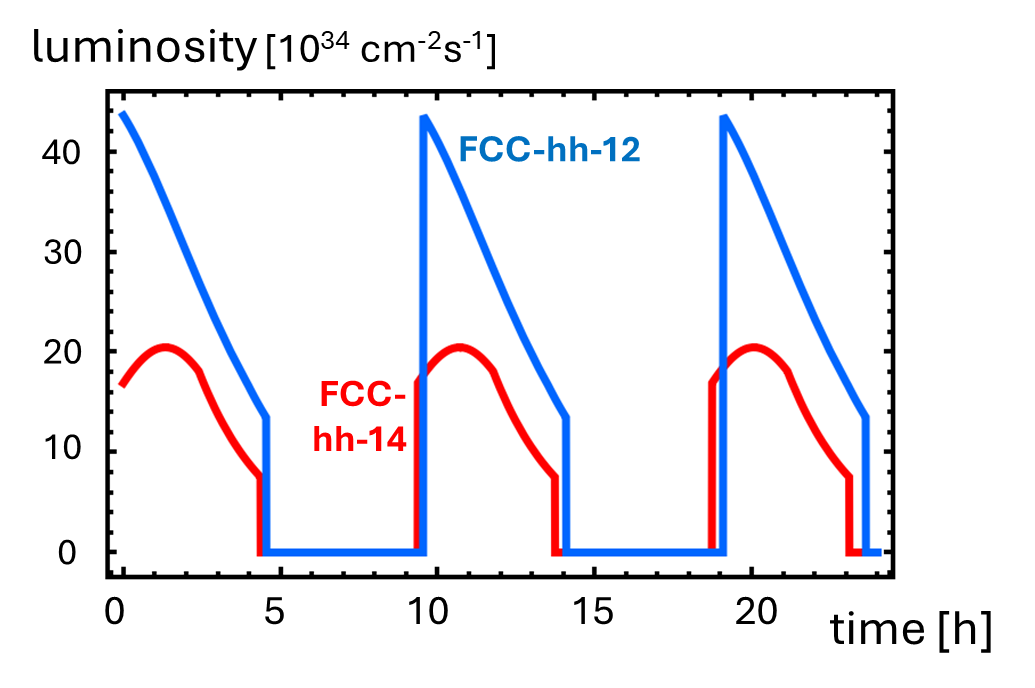}
    \caption{Ideal instantaneous luminosity during 24 h for fixed energy running in scenarios FCC-hh-12 (12 T field) and FCC-hh-14 (14 T dipole field).}
    \label{fig:plumi2}
\end{figure}

\begin{figure}[htb]
    \centering
    \includegraphics[width=\linewidth]{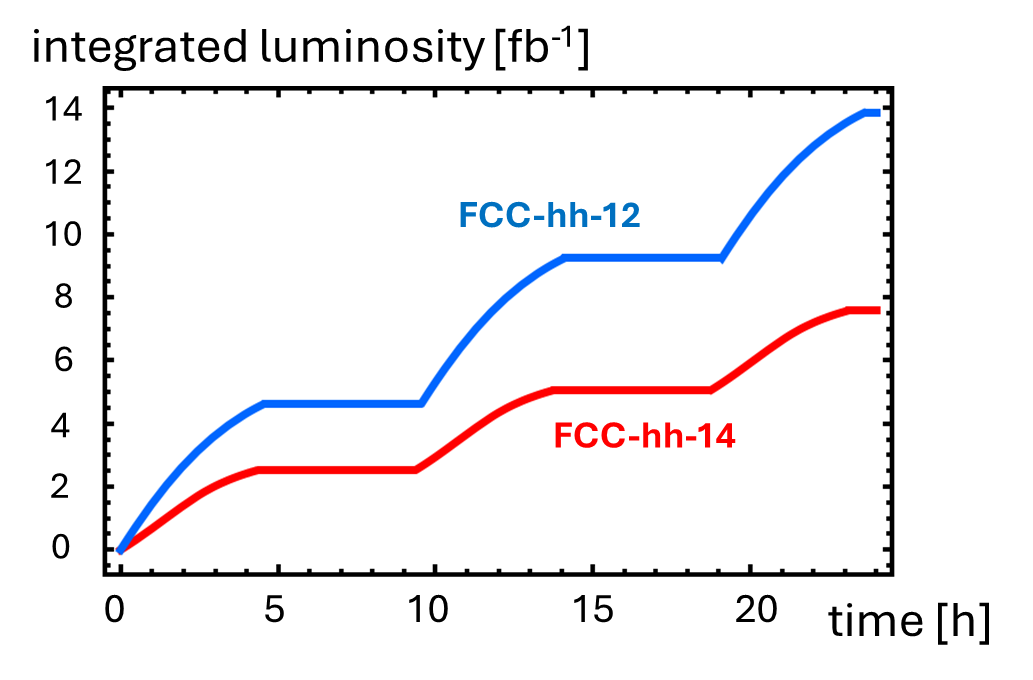}
    \caption{Ideal integrated luminosity during 24 h for fixed energy running in scenarios FCC-hh-12 (12 T field) and FCC-hh-14 (14 T dipole field).}
    \label{fig:pil2}
\end{figure}

For the two scenarios FCC-hh12 and FCC-hh14 from Table \ref{hhfuturetable},
the times $t_1$ are 2.4 h and 4.0 h, respectively.
The optimum run times $t_{r,{\rm opt}}$ turn out to be 
about the same, namely 4.4 h and 4.5 h. 
The time evolution of the instantaneous and integrated luminosity during an ideal day is shown in Figs.~\ref{fig:plumi2} and \ref{fig:pil2}.   
The maximum luminosity is $4.4\times 10^{35}$~cm$^{-2}$s$^{-1}$ and
$2.0\times 10^{35}$~cm$^{-2}$s$^{-1}$ in the two cases. 
The ideal average luminosity for a turnaround time of 
$t_a=5$~hr is $1.3\times 10^{35}$~cm$^{-2}$s$^{-1}$ and
$7.2\times 10^{34}$~cm$^{-2}$s$^{-1}$, respectively.

\section{\label{sec:bd2} One-step leveling dynamics}
The synchrotron radiation power emitted per beam is
\begin{equation}
P_{\rm SR} = f_{\rm rev} n_b N_b \; \frac{ C_{\gamma} E^4}{\rho}  \; ,
\label{srp} 
\end{equation}
where $f_{\rm rev}$ denotes the revolution frequency,
$n_b$ the number of bunches per beam, $N_b$ the bunch population,
$E$ the beam energy, $\rho = (B\rho )/B_{\rm dip}$ the bending radius,
with $(B\ rho)$ designates the magnetic rigidity and $B_{\rm dip}$ 
the arc dipole magnet field, 
and $C_{\gamma} = (4 \pi/3) r_p /(m_p c^2)^3\approx 7.8\times 10^{-18}$~m\; GeV$^{-3}$. 

We first consider a scenario with operation at two different energies.
The running at the first energy level is described by the formulae of Section \ref{sec:bd}. 
According to Eq.~(\ref{srp}), 
the energy could be increased from value $E_1$ to a new level $E_2$, when the bunch intensity has decreased to $N_{b}=N_{0} \left( E_1/E_2 \right)^4$.  
We here consider the case that this happens at a time $t_{2,{\rm step}}$ before $t_1$ from Section \ref{sec:bd}, 
which is fulfilled for our 12 T  example\footnote{In appendix \ref{app:smallert1}, we present formulae for the 
alternative situation that the step time $t_{2,{\rm step}}$ falls 
between the times $t_1$ and $t_2$ from Section \ref{sec:bd}.}.  

Eq.~(\ref{nevol}) yields  
\begin{equation}
t_{2,{\rm step}} = \tau \ln \left( 1 + \left( \frac{E_1^4}{E_2^4} -1  \right) / B \right)   \; .
\end{equation}
The luminosity accumulated at energy $E_1$ is 
\begin{eqnarray}
\Sigma_{1,1} & = & L_0 \tau \left( 
\frac{e^{t_{\rm step}/\tau}-1}{1+B (e^{t_{\rm step}/\tau}-1)}
\right)   \; . 
\label{i1int2}
\end{eqnarray}

After a short ramp time  $\Delta t_{\rm ramp}$, the collisions resume at the higher energy $E_2$. The beam-beam tune shift is unchanged.
Only the time constant $\tau_{2}$ 
of Eq.~(\ref{tau2}) is reduced by the energy ratio 
$E_1/E_2$.  
For times $t>t_{2,{\rm step}} + \Delta t_{\rm ramp}$
the collision occur at the beam energy $E_2$. 
In analogy to (\ref{nevol}), 
the bunch intensity now continues to decrease as
\begin{equation}
N_{b,2} (t) = \frac{N_{2,0}}{B_2  \left( 
\exp (t-t_{2,{\rm step}}-\Delta t_{\rm ramp})/\tau_{E_2}-1 \right) + 1}\; ,
\label{nevol2} 
\end{equation}
where we have introduced 
\begin{equation}
N_{2,0} \equiv \frac{N_0}{B \left(e^{t_{2,{\rm step}} /\tau} -1 \right) +1 }\; ,
\label{n20d}
\end{equation}
\begin{equation}
B_2 \equiv \frac{K \tau_{E,2} N_2}{\varepsilon_0}\, e^{t_{2,{\rm step}}/\tau} \; ,
\label{b2def}
\end{equation}
and $\tau_{E_2} = (E_1/E_2)^3\tau $ denotes the damping 
time at the higher energy $E_2$. 
The behavior changes at the time $t_{1,2}$ when the tune shift reaches the maximum value $\Delta Q_{\rm max}$. 

Introducing the tune shift at the start of the higher-energy 
operation  
\begin{equation}
\Delta Q _{0,2} \equiv n_{\rm IP}  \frac{r_p N_{2,0}}{4\pi \gamma_2 \varepsilon_{0,2} }  \; ,
\label{q02def}
\end{equation}\
with
\begin{equation}
\varepsilon_{0,2}  \simeq \varepsilon_0 \exp \left( -\frac{t_{2,{\rm step }}}{\tau}   \right)\; ,  
\label{emit20}
\end{equation} 
from Eq.~(\ref{dqlim}) we deduce the time limit $t_{1,2}$:
\begin{equation}
t_{1,2} = t_{2,{\rm step}}+\Delta t_{\rm ramp} + \tau_{E_{2}}  \ln 
\left[ \frac{ 
1 - B_2}{\frac{\Delta Q_{0,2}}{\Delta Q_{\rm max}}-B_2
}
\right] \; .
\label{t12}
\end{equation}

In analogy to Eq.~(\ref{i1int}), the integrated luminosity at energy $E_2$ collected till time $t_{1,2} $ reads  
\begin{eqnarray}
\Sigma_{2,1} & = & L_{2,0} /\tau_{E_{2}} \left( \frac{\Delta Q _{\rm max}/ \Delta Q_{2,0} -1}{1-B_2 } \right)   \; , 
\label{i21int}
\end{eqnarray}
with
\begin{equation}
L_{2,0}  \equiv L_0 \; \frac{E_2}{E_1} \; 
\frac{e^{t_{2,\rm step}}{\tau}}{\left(B (
e^{t_{2,{\rm step}}/\tau} - 1 ) +1\right)^2 } 
   \; , 
\label{l20}
\end{equation} 

Following the derivation that led to Eq.~(\ref{ltinst}), 
for times $t>t_{1,2}$
the luminosity decays exponentially 
\begin{equation}
L_{2,2} (t)= L_{2,{\rm max}}  e^{-(t-t_{1,2} )/\tau_{2,2} }
\label{ltinst2}
\end{equation}
with
\begin{equation}
L_{2,{\rm max}}\equiv \frac{n_b }{\sigma_{\rm tot} \tau_{2,2} n_{\rm IP}} 
N_{b,2} (t_{1,2})  
\label{l2max}
\end{equation}
where, in analogy to Eq.~(\ref{tau2}) we have introduced the time constant  
\begin{equation}
\tau_{2,2}  \equiv \frac{\beta^{\ast} r_p}{\sigma_{\rm tot} f_{\rm rev} 
\gamma_2 \Delta Q_{\rm max}} \; 
   \; ,
\label{tau22}
\end{equation} 
with $\gamma_2$ denoting the relativistic 
Lorentz factor at the higher energy ($\gamma_2 \equiv (E_2/E_1)\gamma $).

The total integrated luminosity at time $t>t_{1,2}$ is 
\begin{eqnarray}
\label{i1i22}
\Sigma(t) & =& \Sigma_{1,1}+\Sigma_{2,1} +\int_{t_{1,1}}^t L_{2,2} (t)  dt  \\ 
& = & 
\Sigma_{1,1}+\Sigma_{2,1} 
+ L_{2,{\rm max}} \tau_{2,2} \left( 1 - e^{-\frac{t_r -t_{1,2} }{\tau_{2,2}}} \right) \nonumber
 \; . 
\end{eqnarray}

Specifically, the integrated luminosity for energy $E_1$ 
stays constant at times $t>t_{2,{\rm step}}$, equal to 
\begin{eqnarray}
\label{se1b}
\Sigma_{E_1} & =& \Sigma_{1,1} \nonumber
 \; ,
\end{eqnarray}
while, for $t>t_{2,{\rm step}}+\Delta t_{\rm ramp}$, the integrated luminosity at energy $E_2$ grows as
\begin{eqnarray}
\label{se2b}
\Sigma_{E_2} (t) & =& \Sigma_{2,1} 
+  \\ 
&  & 
 L_{{\rm max},2} \tau_{2,2} 
\left( 1 - e^{
-\frac{t_r-t_{1,2}} {\tau_{2,2}} 
} \right) \nonumber
 \; . 
\end{eqnarray}
To determine the ideal run time $t_{r,2}$, now we can either optimize the total integrated luminosity $(\Sigma_{E_1} + \Sigma_{E_2})$ or the luminosity at  $E_2$, namely $\Sigma_{E_2}$.
Let us consider the second approach, which also  
leads to a higher burn-off of stored protons.

The average luminosity at energy $E_2$ depends on the run time $t_{r,2}$ 
as
\begin{equation}
L_{{\rm ave},E_2} (t_{r,2} )=\frac{\Sigma_{E_2} (t_{r,2} )}{t_{r,2}+t_a }\; .
\label{lavetr2}
\end{equation}

In analogy to Eqs.~(\ref{Ddef}),
we define auxiliary parameters
\begin{eqnarray}
C_2 & \equiv & 
 L_{{\rm max},2} \tau_{2,2} + \Sigma_{2,1} 
\nonumber
\\
D_2 & \equiv & 
- L_{{\rm max},2} \tau_{2,2} 
e^{\frac{t_{1,2} }{\tau_{2,2}}}  \; ,   
\label{D2def}
\end{eqnarray}
and compute the 
optimum run time 
$t_{r,2}$  
 as 
\begin{equation}
D_2 \left( t_{r,2,{\rm opt}}+t_a +\tau_{2,2} \right) 
e^{-t_{r,2,{\rm opt}}/\tau_{2,2} } + \tau_{2,2} C_2 \stackrel{!}{=} 0 \; ,
\label{d2sec4} 
\end{equation}
which can again be solved numerically.

\section{\label{sec:twolev} One-step leveling example}
We start running at a dipole field of 12 T, with the high-intensity beam parameters of FCC-hh-12 in Table \ref{hhfuturetable}.
At time $t_{2,{\rm step}}=2.0$~h, the beam current is so low that we can increase the arc dipole field to 14~T. The maximum beam-beam tune shift of 0.025 has not yet been reached. This would have happened only after about 4 hours.

However, ramping up to 14 T, and  
running  at the high energy, from (\ref{t12}) already  
after a total $t_{1,2}\approx 2.5$~h, the tune shift limit is reached,
and the solution of Eq.~(\ref{d2sec4}) yields an optimum total run time 
of $t_{r,2,{\rm opt}} = 5.8$~h.

Concerning the duration of the field ramp, the FCC-hh full 
ramp time from injection to top energy is of order 20 minutes \cite{AlemanyFernandez:2239138}.
The ramp duration from a dipole field of 12 T to 14 T could then be of order 2 minutes.
Conservatively, including time for adjustments we assume an interval $\Delta t_{\rm ramp}=5$~minutes between the stop of colliding at 12 T and the start of collisions at 14 T. 
These 5 minutes are included in our numbers above.
 
\begin{figure}[htb]
    \centering
    \includegraphics[width=\linewidth]{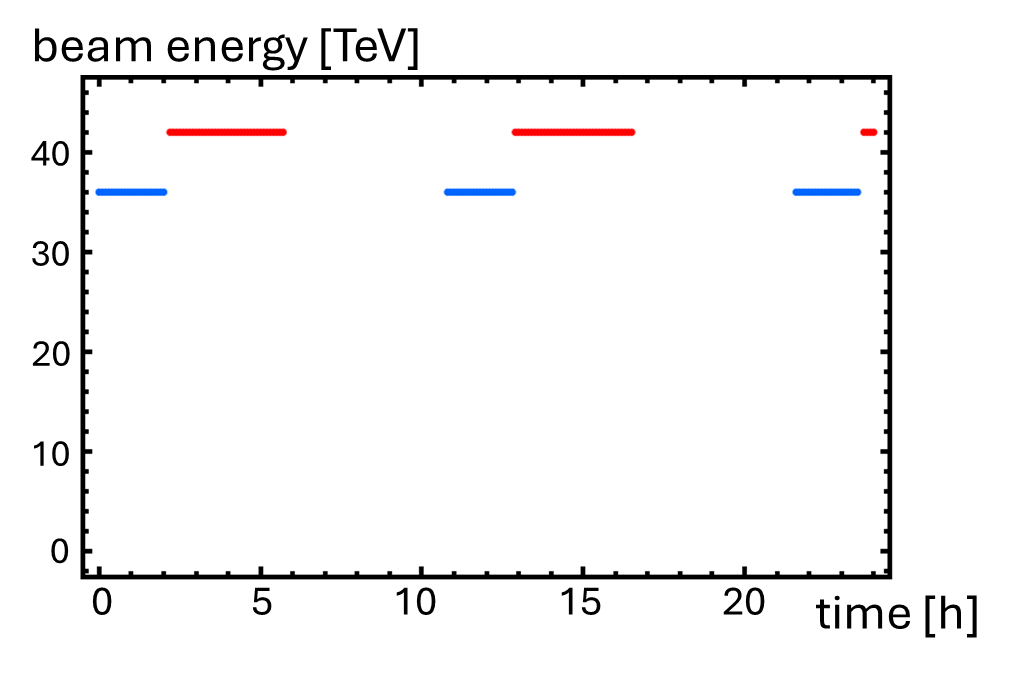}
    \caption{Beam energy in collision during 24 h with one-step leveling, where the dipole field increases from 12 and 14 T about 2.0 h into each physics fill, starting with the parameters of FCC-hh-12.}
    \label{fig:pens}
\end{figure}

Figure \ref{fig:pens} displays the energy of the colliding beams as a function of time.

\begin{figure}[htb]
    \centering
    \includegraphics[width=\linewidth]{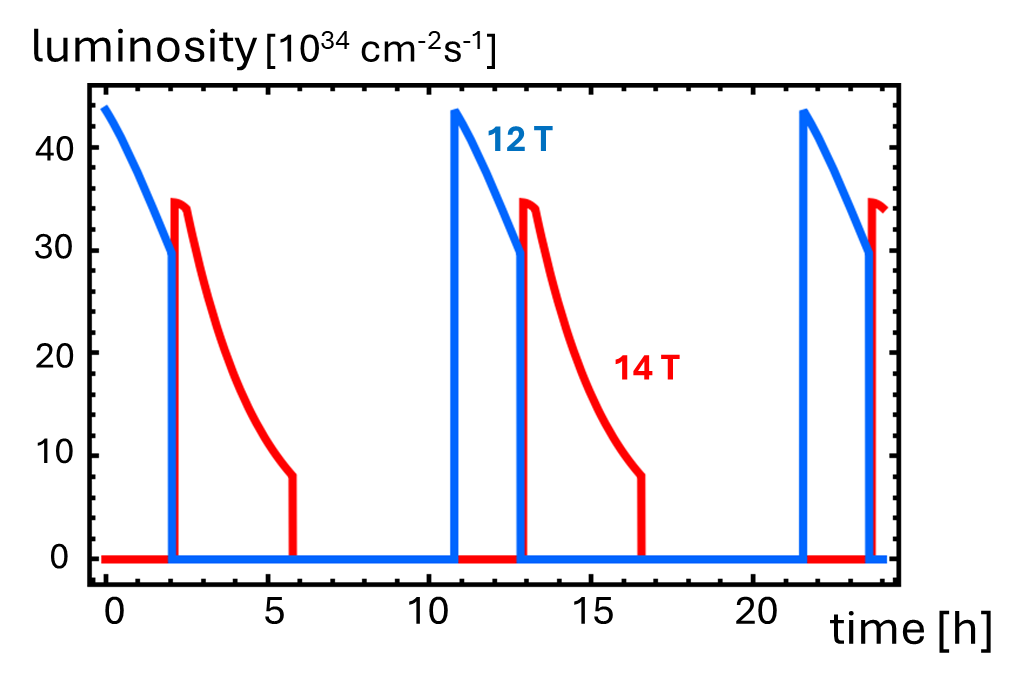}
    \caption{Ideal instantaneous luminosity during 24 h with one-step leveling increasing the dipole field from 12 and 14 T about 2.0 h into each physics fill, starting with the parameters of FCC-hh-12.}
    \label{fig:plumi2lev}
\end{figure}

\begin{figure}[htb]
    \centering
    \includegraphics[width=\linewidth]{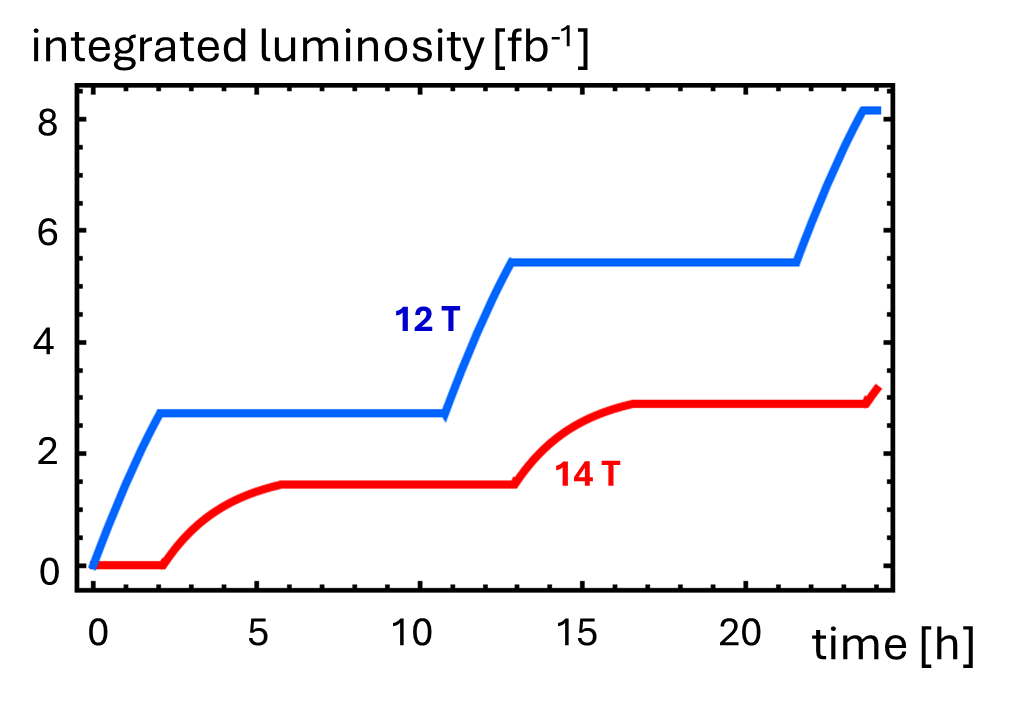}
    \caption{Ideal integrated luminosity during 24 h  with one-step leveling increasing the dipole field from 12 and 14 T about 2.0 h into each physics fill, starting with the parameters of FCC-hh-12..}
    \label{fig:pil2lev}
\end{figure}

\begin{figure}[htb]
    \centering
    \includegraphics[width=\linewidth]{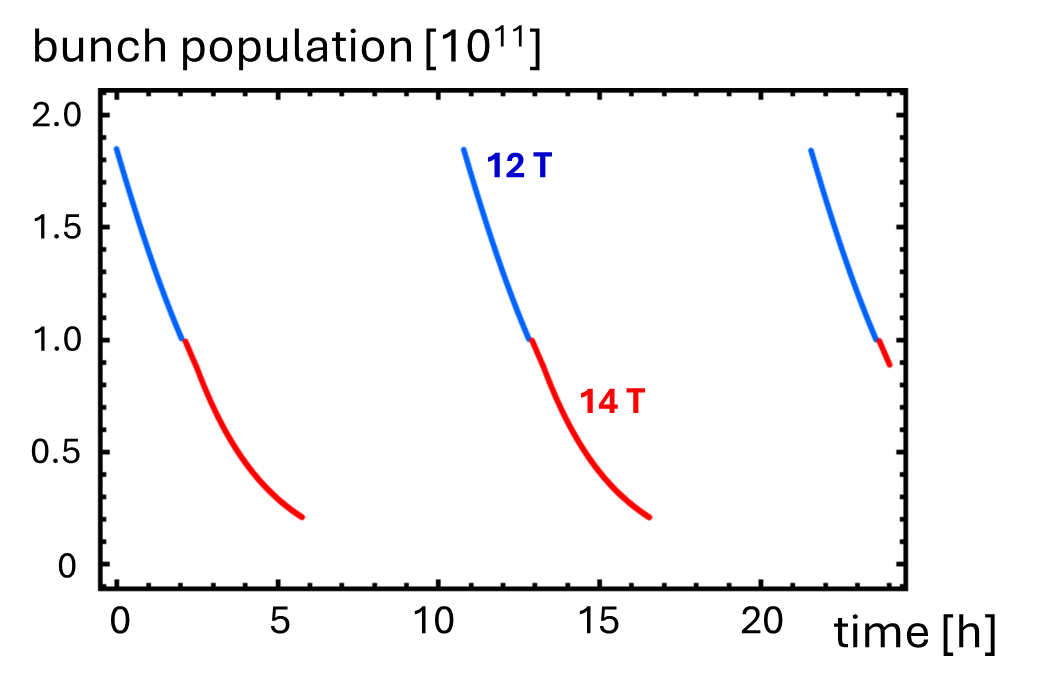}
    \caption{Ideal evolution of bunch intensity during 24 h  with one-step leveling increasing the dipole field from 12 and 14 T about 2.0 h into each physics fill, starting with the parameters of FCC-hh-12..}
    \label{fig:nb2lev}
\end{figure}

The time evolution of the instantaneous and integrated luminosity during an ideal 
day with one-step leveling at dipole fields of 12 and 14 T
is shown in Figs.~\ref{fig:plumi2lev} and \ref{fig:pil2lev}.   
For completeness, in Fig.~\ref{fig:nb2lev}, we present the 
corresponding evolution of the bunch intensity.  

The maximum luminosity is $4.4\times 10^{35}$~cm$^{-2}$s$^{-1}$ and
$3.5\times 10^{35}$~cm$^{-2}$s$^{-1}$ at the 
two energies, respectively, which at 12 T is the 
same peak value as when operating only at that single 
energy, but at 14 T is higher than for the 
exclusive operation at 14 T---compare, e.g., Figs.~\ref{fig:plumi2lev} and \ref{fig:plumi2}.  

In this case, the ideal average luminosities for a 
turnaround time of 
$t_a=5$~hr are $7.0\times 10^{34}$~cm$^{-2}$s$^{-1}$ 
at 72.5 TeV (12 T) and 
$6.7\times 10^{34}$~cm$^{-2}$s$^{-1}$  
at 84.6 TeV (14 T field), respectively,
which are being accumulated together.

\section{\label{sec:cont} Continuous leveling}
In this section we aim at maintaining a constant synchrotron radiation power throughout a physics fill, by continually adjusting the beam energy. 
From Eq.~(\ref{srp}), a constant synchrotron radiation power implies
an increase in beam energy $E$, 
as the bunch population $N_b$ decreases, according to
\begin{equation}
    \frac{dE}{dt} = -\frac{1}{4} \frac{E(t)}{N_b(t)}
    \frac{dN_{b}}{dt}\; ,
    \label{cont1}
\end{equation}
From Eq.~(\ref{cont1}), we deduce that 
\begin{equation}
    \frac{dE}{E} = -\frac{1}{4} \frac{dN_{b}}{N_b}\; ,
\end{equation}
leading to the relation 
\begin{equation}
 E(t) =  E_0 \left ( \frac{N_0}{N_b(t)} \right)^{1/4}\; . 
 \label{enbrel}
\end{equation}

On the other hand, for $dN_b/dt$ in (\ref{cont1}) 
we can insert the expression (\ref{dnbdt}) to obtain
\begin{equation}
    \frac{dE}{dt} = \frac{K}{4} \frac{E(t) N_b(t)}{\varepsilon (t)}\; . 
    \label{dedt}
\end{equation}


For simplicity, here we ignore the initial period 
where the beam-beam tune shift still increases due to radiation damping. 
The intensity and energy variation in this period is 
discussed in Appendix \ref{app:ueq}. 
Instead we assume that the 
starting beam-beam parameters are already 
at the beam-beam limit. 
We can vary the value of this assumed limit in order 
to indicate the possible range of luminosity performance. 

At the beam-beam limit the total tune shift $\Delta Q_{\rm max}$ is constant,
constraining the ratio $N_b(t)/\varepsilon(t)$ as 
\begin{equation} 
\frac{N_b (t) }{\varepsilon (t) } = \frac{ \Delta Q_{\rm max} 4 \pi E(t)}{r_p n_{\rm IP} m_ec^2}
\label{nberel}
\end{equation}
Inserting this into (\ref{dedt}) and also using the
definition of $K$ from (\ref{kdef}), we obtain
\begin{equation}
    \frac{dE}{dt} = E(t)^2 \; F \; ,
    \label{dedt3}
\end{equation}
with
\begin{equation}
F\equiv  \frac{ \Delta Q_{\rm max} \sigma_{\rm tot} f_{\rm rev}} 
    {4 \beta^{\ast} r_p m_e c^2 }\; .
\label{fdef} 
\end{equation}
Integration yields 
\begin{equation}
    E(t)  = E_0 \; \frac{1}{1- F E_0 t} \; ,
    \label{ec2}
\end{equation}
which is accompanied by the change in bunch intensity 
\begin{equation}
    N_b (t)  = N_0 \; \left( 1- F E_0 t \right)^4 \; .
    \label{nbc2}
\end{equation}
Combining Eqs.~(\ref{nberel}), (\ref{nbc2}) 
and (\ref{lumi}), we can write the luminosity as

\begin{eqnarray}
  L (t) &  = & \frac{f_{\rm rev} n_b }{\beta^{\ast} } 
  \frac{ \Delta Q_{\rm max} }{r_p n_{\rm IP} m_ec^2}
   \; E(t)\; N_b(t) \nonumber \\
    & = & 
G  \; E_0 N_0 \;  \left( 1- F E_0 t \right)^3 \; ,
    \label{lc2}
\end{eqnarray} 
with
\begin{equation} 
    G\equiv \frac{f_{\rm rev} n_b \Delta Q_{\rm max}}{r_p n_{\rm IP} \beta^{\ast} m_ec^2}\; '
    \end{equation} 
The integrated luminosity after tine $t_l$  reads
\begin{equation} 
    \Sigma_{\rm cont} (t_l)  \equiv \int _{0}^{t_l} L(t)\; dt = 
    \frac{GN_0}{4 F} \left[ 1- \left(1 - F E_0 t_l\right)^4\right] \; .
    \label{scont}
    \end{equation}

From (\ref{ec2}), for a given maximum beam energy $E_{\rm max}$, 
the leveling lasts for a time period $t_{\rm lev}$, 
\begin{equation}
t_{\rm lev} = \frac{1}{F} \left( \frac{1}{E_0}-\frac{1}{E}\right) \; .  
\label{eq:tropt} 
\end{equation}

 The differential integrated 
 luminosity spectrum during one leveling period 
 is found from the relation 
 $d \Sigma_{\rm cont}/dE = |d \Sigma_{\rm cont}/dt| / (dE/dt )$. 
 Using Eqs.~(\ref{ec2}) and (\ref{scont}), this yields
 \begin{equation} 
\frac{d\Sigma_{\rm cont}}{dE} = \frac{ G N_0}{FE_0} \left( \frac{E_0}{E} \right)^5\; ,  
\label{dsde}
    \end{equation}
which extends from $E=E_0$ to $E_{\rm max} = E_0/(1 - FE_0 t_{\rm lev})$.

From this moment in time onward, 
the collider can continue to run at the constant maximum energy.
In analogy to (\ref{n2evol2}), the bunch intensity 
now decreases as 
\begin{equation}
N(t) =N_{2}e^{-(t-t_{\rm lev})/\tau_{2,c} }\; , 
\label{n2evol2c} 
\end{equation}
with 
\begin{eqnarray}
N_{2} & \equiv & N_0\, \left( 1 - F E_0 t_{\rm lev}  \right) ^{4}  
 \; \\
\tau_{2,c} & \equiv &  \frac{\beta^{\ast} r_p}{\sigma_{\rm tot} f_{\rm rev} \gamma_{\rm max} \Delta Q_{\rm max}} \; ,
\end{eqnarray}
and the luminosity as 
\begin{equation}
L_2 (t)= L_{{\rm max},c}  e^{-(t-t_{\rm lev}  )/\tau_{2,c} }
\label{ltinstc}
\end{equation}
where 
\begin{equation}
L_{{\rm max},c}\equiv  \,
G E_0 N_0\, \left( 1 - F E_0 t_{\rm lev}  \right) ^{3} 
 \; .
\label{lmaxc}
\end{equation}
The total integrated luminosity after time $t>t_{\rm lev}$ is
\begin{eqnarray}
\label{sec}
\Sigma_{\rm tot} (t) & =& \Sigma_{\rm cont} (t_{\rm lev})
+ L_{{\rm max},c} \tau_{2,c} \left( 1 - e^{-\frac{t -t_{\rm lev} }{\tau_{2,c}}} \right) 
 \; ,
\end{eqnarray}
where $\Sigma_{\rm cont}$ was given in (\ref{scont}). 
As before, we maximize the average luminosity 
$\Sigma_{\rm cont}/(t_{ta}+t_r)$. 
The corresponding condition   
\begin{equation} 
   \frac{d}{dt_r} \left( \frac{ \Sigma_{\rm tot} (t_r)}{t_{\rm ta}+ t_r} \right) _{t_r=t_{r,{\rm op}}} 
   \stackrel{!}{=} 0 \; ,
    \end{equation}
yields the optimum
run time $t_r=t_{r,{\rm opt}}$ as in 
Eq.~(\ref{ort1}): 
\begin{equation} 
 D_c \left( t_{r,{\rm opt}}+t_a +\tau_{2,c} \right) 
e^{-t_{r,{\rm opt}}/\tau_{2,c} } + \tau_{2,c} C_c \stackrel{!}{=} 0 \; ,
\label{ort2} 
\end{equation}
where now we have 
\begin{eqnarray}
C_c & \equiv & 
 L_{{\rm max},c} \tau_{2,c} +
     \frac{GN_0}{4F} \left[ 1- \left(1 - F E_0 t_{\rm lev} \right)^4\right]
\nonumber
\\
D_c & \equiv & 
- L_{{\rm max},c} \tau_{2,c} 
e^{\frac{t_{\rm lev}}{\tau_{2,c}}}  \; .   
\label{Dcdef}
\end{eqnarray}

\section{\label{sec:contlev} Continuous leveling example}
\begin{figure}[htb]
    \centering
\includegraphics[width=\linewidth]{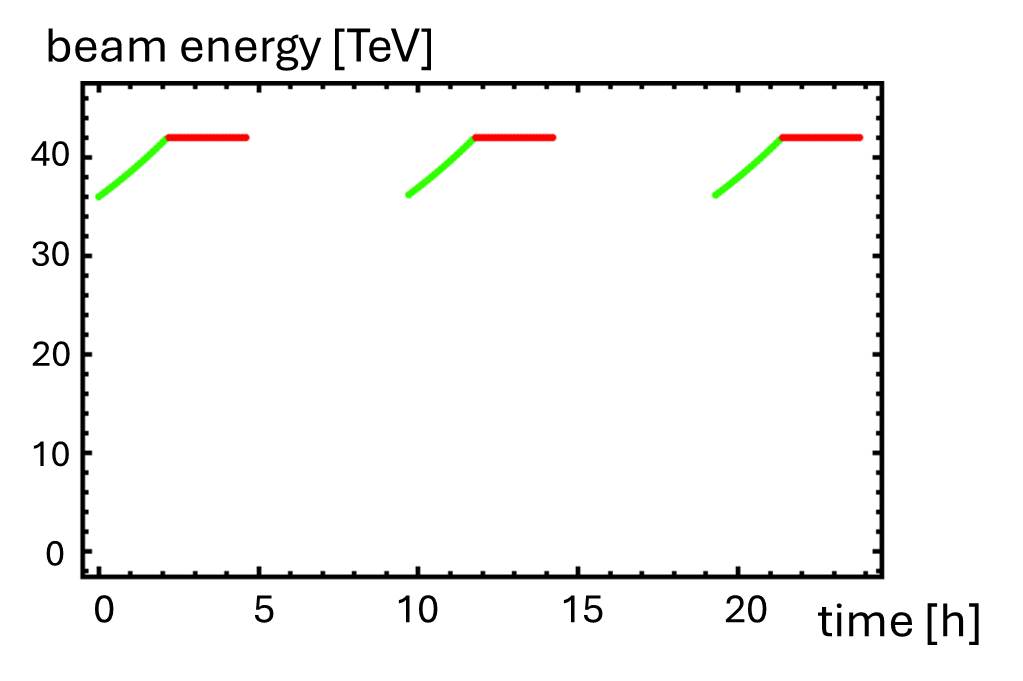}
    \caption{Beam energy in collision during 24 h for a continuous leveling
    at a constant total beam-beam tune shift of $\Delta Q_{\rm tot}=0.0177$, 
    with arc dipole magnet field increasing from 12 to 14 T (green), followed by 
    a final period at constant 14 T field (red).}
    \label{fig:penc}
\end{figure}

We consider a continuous leveling starting at a dipole magnet field of 12 T
or a beam energy of 36 TeV, and ending at 14 T or 42 TeV, with 
otherwise the same initial parameters as for the 12 T case 
in the previous sections,
so that the total beam-beam tune shift of Eq.~(\ref{eq:dQtot}) 
is held constant at about 
$\Delta Q_{\rm tot} = \Delta Q_{0}\approx 0.0177$. 
In this case, 
the leveling stops after $t_{\rm lev}\approx 2.15$~h, 
as is illustrated by the energy profile in  
Fig.~\ref{fig:penc}. 
Almost the same amount of 
luminosity is accumulated at the final 
constant energy of 42 TeV.
The optimum total run time is found to be 
$t_{r,{\rm opt}}\approx 4.6$~h and, hence, more than 
twice the continuous leveling time $t_{\rm lev}$.

\begin{figure}[htb]
    \centering
\includegraphics[width=\linewidth]{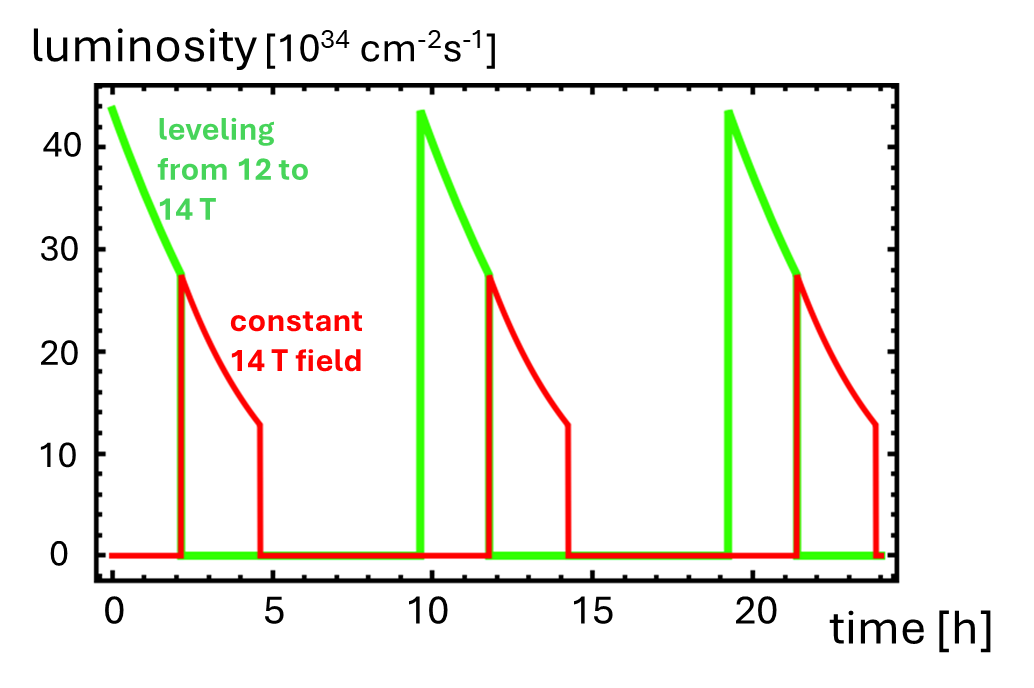}
    \caption{Ideal instantaneous luminosity during 24 h for a  continuous leveling
    at a constant total beam-beam tune shift of $\Delta Q_{\rm tot}=0.0177$, 
    with arc dipole magnet field increasing from 12 to 14 T (green), followed by 
    a final period at constant 14 T field (red).}
    \label{fig:plumic}
\end{figure}

\begin{figure}[htb]
    \centering
    \includegraphics[width=\linewidth]{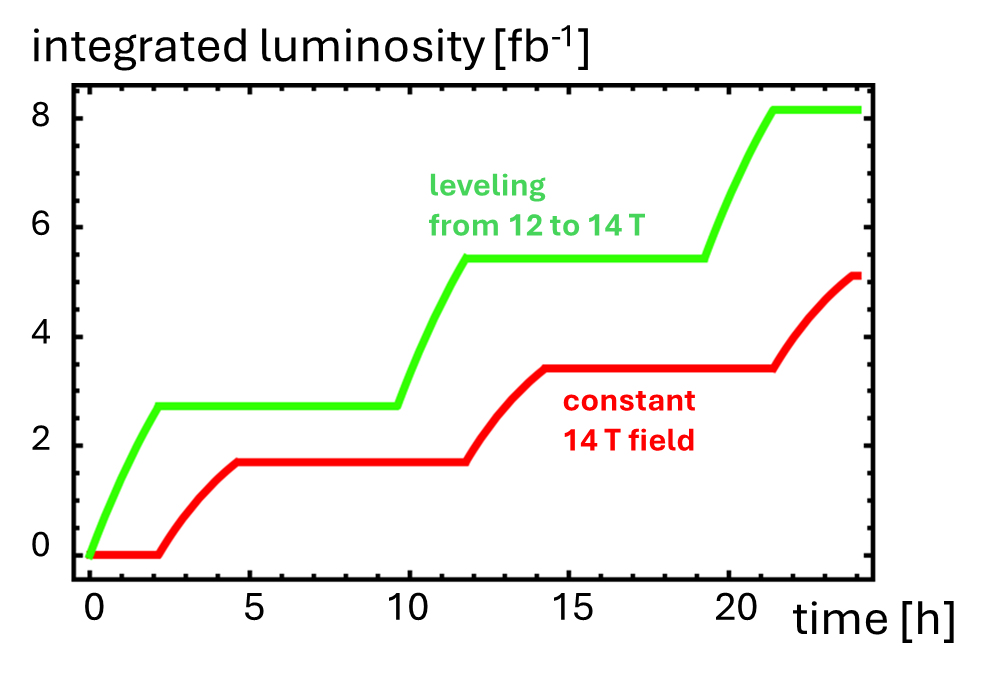}
    \caption{Ideal integrated luminosity during 24 h for a  continuous leveling
    at a constant total beam-beam tune shift of $\Delta Q_{\rm tot}=0.0177$, 
    with arc dipole magnet field increasing from 12 to 14 T (green), followed by 
    a final period at constant 14 T field (red).}
    \label{fig:pilc}
\end{figure}

The time evolutions of the instantaneous and integrated 
luminosity during an ideal day are 
shown in Figs.~\ref{fig:plumic} and \ref{fig:pilc}.   
The ideal average luminosity 
$\Sigma_{\rm cont}/(t_{ta}+t_r)$ amounts to 
$1.3\times 10^{35}$~cm$^{-2}$s$^{-1}$, of which
$7.9\times 10^{34}$~cm$^{-2}$s$^{-1}$ are produced 
during the leveling till $t=t_{\rm lev}$  
and $4.9\times 10^{34}$~cm$^{-2}$s$^{-1}$ 
in the final running with maximum energy at 14 T field.

\begin{figure}[htb]
    \centering
\includegraphics[width=\linewidth]{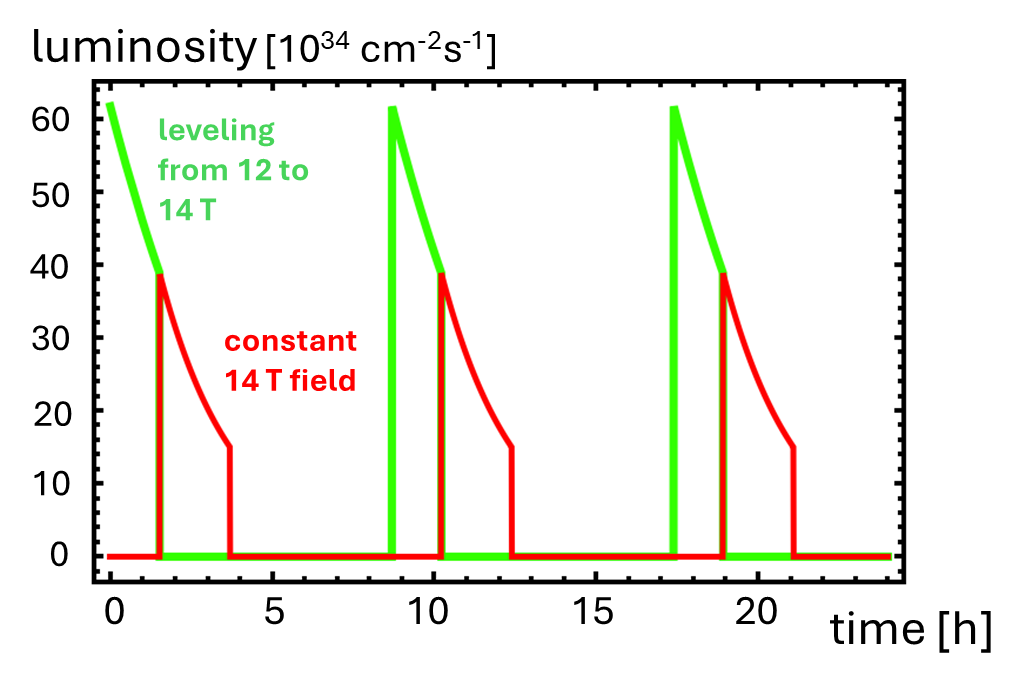}
    \caption{Ideal instantaneous luminosity during 24 h for a  continuous leveling
    at a higher 
    constant total beam-beam tune shift of $\Delta Q_{\rm tot}=0.025$, 
    with arc dipole magnet field increasing from 12 to 14 T (green), followed by 
    a final period at constant 14 T field (red).}
    \label{fig:plumic25}
\end{figure}

For a higher constant beam-beam tune shift of 
$\Delta Q_{\rm tot}=0.025$, instead of 0.0177, 
the instantaneous
luminosity increases by about 40\% and the physics fills are about 30\% shorter,  
as is illustrated in Fig.~\ref{fig:plumic25},
to be compared with 
Fig.~\ref{fig:plumic}.

\begin{figure}[htb]
    \centering
    \includegraphics[width=\linewidth]{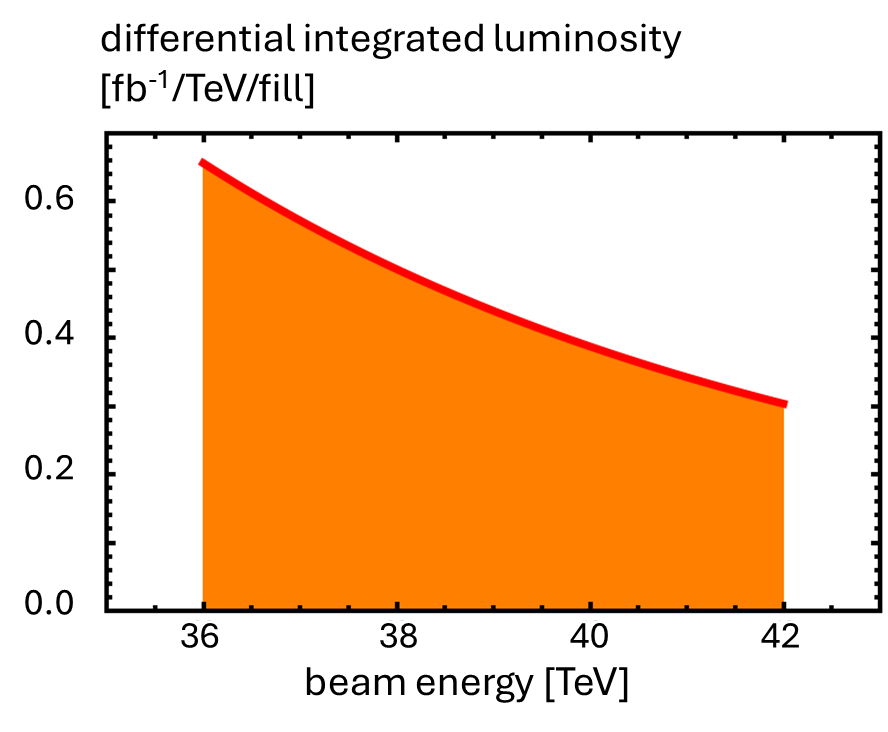}
    \caption{Differential integrated luminosity spectrum for a single continuous leveling period 
    of length  $t_{\rm lev}\approx 1.52$~h, 
    extending 
    from 36 to 42 TeV beam energy, corresponding
    to the first green run portion 
    in Figs.~\ref{fig:penc} and \ref{fig:plumic}.}
    \label{fig:dsde}
\end{figure}

Figure \ref{fig:dsde} presents the differential integrated 
luminosity spectrum accumulated during the 
continuous leveling period of a single fill,
according to Eq.~(\ref{dsde}).
This figure is independent of the assumed
beam-beam tune shift. 

\begin{figure}[htb]
    \centering
     \includegraphics[width=\linewidth]{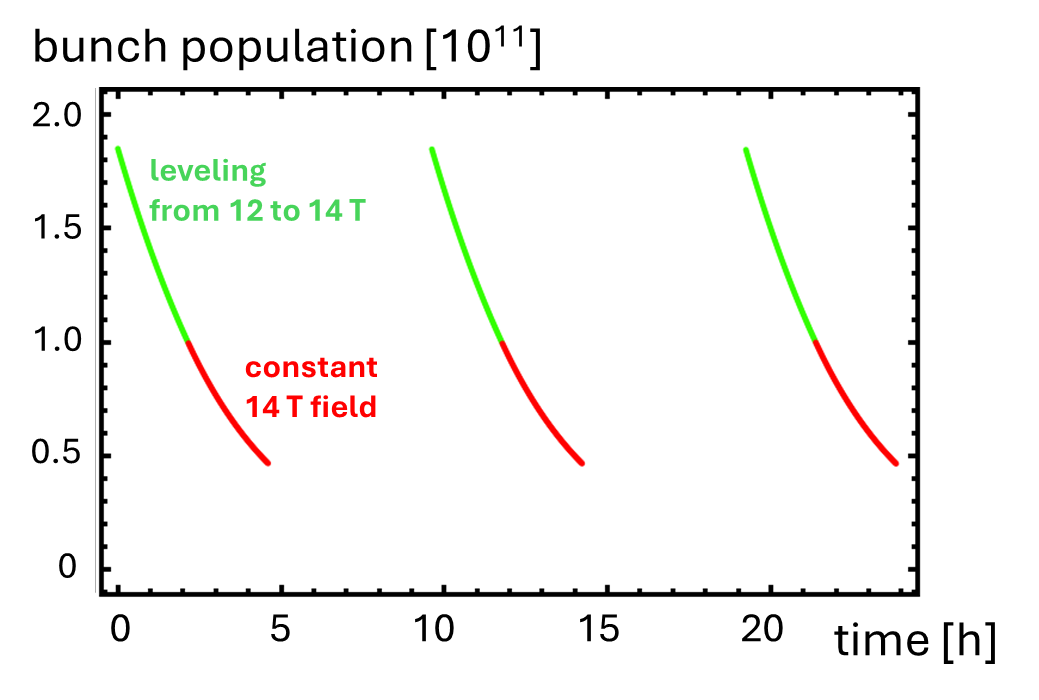}
    \caption{Ideal evolution of bunch intensity during 24 h for a  continuous leveling
    at a constant total beam-beam tune shift of $\Delta Q_{\rm max}=0.0177$, 
    with arc dipole magnet field increasing from 12 to 14 T (green), followed by 
    a final period at constant 14 T field (red).}
    \label{fig:nbc}
\end{figure}

\section{\label{sec:disc} Discussion}
The projected performance of the different scenarios 
is summarized in Table \ref{disctable}.

For the continuous leveling, two examples are listed.
The first is the one considered in the previous section with a constant
total beam-beam tune shift of $\Delta Q_{\rm tot} =0.0177$.
The second (last column) represents the continuous 
leveling at the higher constant tune shift  of 
$\Delta Q_{\rm tot} =0.025$, which corresponds 
to the maximum value of the one-step 
leveling case and for the two fixed-energy running examples. 
This leads to an initial luminosity of 
$6.1\times 10^{35}$~cm$^{-2}$s$^{-1}$ 
(Fig.~\ref{fig:plumic25}).  

The various leveling schemes accomplish roughly 
the  same total integrated luminosity of 
about 1.3~ab$^{-1}$ per year per experiment, as the
operation at the constant lower field of 12 T. 
The advantage is that more than a third of this
integrated luminosity, namely 
0.5--0.7~ab$^{-1}$ per year is delivered at 14 T.
The continuous leveling produces another $\sim$1.0~ab$^{-1}$
at intermediate fields or collision 
energies between 72 and 84 TeV, with 
a distribution as shown in Fig.~\ref{fig:dsde}.
Operation at a constant field of 14 T would only yield
a total luminosity of 0.7~ab$^{-1}$ per year.
In terms of overall performance the continuous leveling 
scheme appears to be best, with the one-step leveling 
scheme not too far behind, and with, probably, more
relaxed implications for the experiments,
as discussed next.

\begin{table*}[htbp]
\caption{Optimized run parameters of FCC-hh 
at 12 T and 14 T dipole field (or in between),  
corresponding to proton-proton 
centre-of-mass collision energies of 
(or between) 72 and 84 TeV, respectively,  
for constant energy operation,
one step energy leveling, and continuous energy leveling.
The integrated luminosity per year and interaction point is estimated by considering 
a total time of 160 days per year allocated for physics running and operation with 70\% efficiency, i.e., 
taking 70\% of the ideally attainable 
luminosity, that would be obtained when repeating the optimum run cycles derived in this report without any interruptions or failures. 
The di-Higgs production rate is discussed in Section \ref{sec:det}.  
}
\label{disctable}

\centering 
\begin{tabular}{lcccccc}
\hline\hline
 & unit & 12\,T & 14\,T & one step & continuous & continuous \\
  &  & fixed & fixed & leveling & leveling & leveling \\
\hline
initial total beam-beam tune shift $\Delta Q_{{\rm tot},0}$ & & 0.0177, & 0.011 & 0.0177 & 0.0177 & 0.025\\ 
maximum total beam-beam tne shift $\Delta Q_{\rm tot,\, max}$ & & 0.025 &  0.025 & 0.025 & 0.0177 & 0.025\\ 
peak luminosity at 12 T  & nb$^{-1}$s$^{-1}$ & 440 & --- & 440  & 440 & 610 \\
peak luminosity at 14 T  & nb$^{-1}$s$^{-1}$ &  --- & 200 & 350  & 275 & 380 \\
run time at 12 T  &  h  &  4.4 & ---   & 2.5  & N/A & N/A  \\
run time at intermediate fields  between
12 and 14 T  &  h  &  --- & ---   & ---  & 2.1 & 1.5 \\
run time at 14 T  &   h & ---  & 4.5  &  3.2  & 2.5 & 2.2 \\
total fill length  & h & 4.4  & 4.5 & 5.8  &  4.6 & 3.7 \\
time-averaged ~luminosity at 12 T & nb$^{-1}$s$^{-1}$ & 135  &  --- &  70 & --- & ---\\
time-averaged~luminosity at fields between 12 and 14 T\; \; \;  & nb$^{-1}$s$^{-1}$ & --- & --- &  --- & 79  &  87  \\
time-averaged luminosity at 14 T & nb$^{-1}$s$^{-1}$ & --- & 75 &  67 & 49 &  63\\
integrated ~luminosity at 12 T per interaction point & ab$^{-1}$/y & 1.3 & --- & 0.7 & --- & --- \\
integrated ~luminosity at 12--14 T per interaction point & ab$^{-1}$/y & --- & --- & --- & 0.8 & 1.0 \\
integrated ~luminosity at 14 T per interaction point & ab$^{-1}$/y & --- & 0.7 & 0.6 & 0.5 & 0.7 \\
total~integrated ~luminosity  per interaction point& ab$^{-1}$/y & 1.3 & 0.7 & 1.3 & 1.3 & 1.7 \\
unburnt fraction of protons $N_b (t_{\rm r})/N_0$ & \% &  22 &  19  &  11  & 28 & 21 \\
di-Higgs production rate normalized 
to 14 T fixed field  & \% & 145 &  100 & 164 &  157 & 220 \\
\hline\hline
\end{tabular}
\end{table*}

\section{Detector \& Physics Considerations}
\label{sec:det}
In the same way as the LHC experiments run with different pileup conditions, the FCC-hh detectors could run at different collision energies.   

The synchrotron radiation power leveling offers 60\% to 120\% 
more luminosity than running at a constant energy with a fixed magnetic 
field of 14 T. 
To understand whether this is helpful the energy dependence of the 
relevant cross sections must be taken into account. 

The di-Higgs production is a key physics deliverable, which can be obtained already  
at rather low centre-of-mass energy~\cite{cavaliere2025physicsprospectsneartermprotonproton}. 
For this example process, the luminosity can be  
more important than the energy.

Using a quadratic fit as in Ref.~\cite{cavaliere2025physicsprospectsneartermprotonproton}, 
to extrapolate the di-Higgs cross section as a function of the centre-of-mass energy
we obtain 
\begin{eqnarray}
    \lefteqn{\sigma_{\rm di-Higgs}\, [{\rm fb}] 
    \approx }   \label{eq:dihiggs} \\
& &     0.0352 e^{-0.1646 (\log E_{\rm cm}\, [{\rm TeV}]) ^2} (\log E_{\rm cm}\, [{\rm TeV}])^{3.061} \nonumber 
\end{eqnarray}
The cross section for di-Higgs production is about 
29\%  higher at 84 TeV than at 72 TeV, or 1077~fb compared with 838~fb. 

From these numbers together with Table \ref{disctable}, 
we can estimate that with one-step levelling the FCC-hh would produce about 64\%
more di-Higgs events than with only 14 T field operation, and still
about 13\% more di-Higgs events than with only 12 T field operation.

To infer the average di-Higgs production rate 
with continuous leveling, 
we integrate in energy 
over the product of the differential luminosity spectrum of 
Fig.~\ref{fig:dsde} and the cross section of Eq.~(\ref{eq:dihiggs}).
Depending on the beam-beam tune shift value considered, the result is a di-Higgs production rate 
that is either about the same as, or up to 35\% higher than for the one-step leveling.
This means that the continuous leveling could yield up to 2.2 times 
more di-Higgs events than for operation at a constant dipole field of 14 T.  

For the experiments, a stepwise leveling appears  feasible  \cite{hoecker}.
It would also be easier than the continuous leveling \cite{wengler},  
as all cross sections are energy-dependent. 
The experiments could simply discard data from 
the energy ramp (if any)  and split each run into 2 (or more)  datasets \cite{gray}. 
The experimental trigger menus could be adjusted 
via a `key' for each step, so as to make sure that the event selection always remains optimum \cite{hoecker}. The Monte-Carlo (MC) 
productions should then simulate the same scenarios, 
including event generation. 
As a positive side-effect, identical detector conditions at different 
energies would, possibly for the first time, 
enable high-precision ratio measurements \cite{gray}.

On the other hand, 
continuous leveling would imply major 
changes in the way 
simulations and data analysis are handled \cite{gray}. 
The amount of additional data would need to be significant 
to make tackling the associated challenges worthwhile.  
A continuously changing beam energy would require a finite number of energy steps in simulation with some interpolation algorithms in between \cite{wengler}. 
Such approach was, indeed, taken at the end of LEP2 with 
the ``mini-ramps'' in energy  \cite{wengler,miniramp} 
--- but there the energy steps were quite small and the interpolation not very problematic \cite{wengler}. 

Present MC generators for the LHC experiments  
cannot handle varying centre-of-mass energies.  
For the continuous leveling scenario, a software development would be needed, taking into account changes in the cross section, phase space, and so on \cite{gomez}. One could then re-weight the simulated samples as a function of both pileup and centre-of-mass energy \cite{gomez}. Also the proton parton distribution functions (PDFs) change as a function of energy.
Precision cross section measurements would become more difficult too,
and the energy dependence of the cross sections would render 
comparison to theoretical predictions more challenging.

For all these reasons, the one-step leveling or a staircase of 3 or 4 discrete steps would be easier than continuous leveling, and could be simulated directly without interpolations \cite{gray}.
A few-step leveling scheme would come close to the performance of the continuous leveling. 
However, as the number of steps increases, the luminosity per step decreases and, hence, 
so does the precision of the measurement at each energy point, 
even though the total precision of the combination increases with the additional luminosity.

In general, if synchrotron-radiation leveling at FCC-hh is accompanied by a significant physics gain, support from the experiments is probable \cite{hoecker}.

\section{\label{sec:cons}Conclusions}
Synchrotron radiation power leveling is a possible
novel operation mode for future highest-energy hadron colliders  whose beam currents  are limited by the cryogenic power, such as the proposed 
future circular hadron collider FCC-hh.  

In this new mode of operation, during each physics fill, the arc magnetic dipole field and, thereby, the collision energy is increased, either in steps or continually, in order to maximize the total integrated luminosity, while extending the exploration and physics reach 
up to maximum reachable fields and energies. 

In this article, we have derived analytical formulae describing the time dependence of beam current, luminosity and energy, and the optimum physics fill length, considering different scenarios. 
Pictures illustrate the various running modes. 

The two leveling scenarios --- stepwise or continuous ---
have their respective advantages. One would be easier  to implement; 
the other would yield even higher average event rates.  
To realize either of them, 
the experimental detectors would need to be 
prepared for this new approach to colliding-beam operation, 
which is expected to be much simpler 
for the case of one- (or few-) step leveling. 

The predicted significant increases in certain key 
production rates, e.g., a 64\% or up to 120\% rise 
in the number of di-Higgs events, for either leveling 
scenario, respectively, 
would justify such new modes of operation.

\section*{\label{sec:ack}Acknowledgements}
I most warmly thank E.~Lipeles, who, while making the case for higher luminosity at lower FCC-hh 
collision energy, suggested this novel mode of hadron collider operation \cite{lipeles}.
I would also like to extend my sincere thanks to 
A.~Canepa, G.~Gomez-Ceballos, H.~Gray, A.~H\"{o}cker, and T.~Wengler for extremely 
helpful discussions on the associated detector challenges. 
Finally, I gratefully acknowledge K.~Jakobs for his unwavering commitment to future hadron colliders and for his steadfast support.

\bibliography{bibliography} 

\appendix

\section{Two-step leveling with step at the beam-beam limit}
\label{app:smallert1} 
In Section \ref{sec:bd2} we considered two-step leveling under the assumption 
that the beam-beam limit was not yet reached at the lower energy, e.g., for $t_{2,{\rm step}}<t_1$. 
Here, we present formulae corresponding to 
the case that  
that the step time $t_{2,{\rm step}}$ 
falls between the times $t_1$ and $t_2$ 
from Section \ref{sec:bd}, 
This situation would occur, e.g., at a higher value for the initial  
beam energy or with a smaller initial emittance. 

Equation~(\ref{n2evol}) yields  
\begin{equation}
t_{2,{\rm step}} = t_1 - 4 \tau_2 \ln \left( \frac{E_1}{E_2} \right) -\tau_2 \ln \left( B
\left( e^{\frac{t_1}{\tau}} -1 \right)+1\right) \; .
\end{equation}

After a short ramp time  $\Delta t_{\rm ramp}$, the collisions resume at the higher energy $E_2$. The beam-beam tune shift is unchanged.
Only the time constant $\tau_{2}$ 
of Eq.~(\ref{tau2}) is reduced by the energy ratio 
$E_1/E_2$.  
For times $t>t_{2,{\rm step}} + \Delta t_{\rm ramp}$
the collision occur at the beam energy $E_2$. 
In analogy to (\ref{n2evol}) and (\ref{ltinst}), the bunch intensity 
now decreases as 
\begin{equation}
N(t) =N_{2,0}e^{-(t-t_{2,{\rm step}}-\Delta t_{\rm ramp} )/\tau_{2,2} }\; , 
\label{n2evol2} 
\end{equation}
with 
\begin{eqnarray}
N_{2,0} & \equiv & N_0\,  
\frac{e^{-\frac{t_{2,{\rm step}}-t_1}{\tau_{2}}}}
{B ( e^{t_1/\tau_1}-1 )+ 1} \\
\tau_{2,2} & \equiv &  \frac{\beta^{\ast} r_p}{\sigma_{\rm tot} f_{\rm rev} \gamma_2 \Delta Q_{\rm max}} \; ,
\end{eqnarray}
and the luminosity as 
\begin{equation}
L_2 (t)= L_{{\rm max},2}  e^{-(t-t_{2,{\rm step}} -\Delta t_{\rm ramp} )/\tau_{2,2} }
\label{ltinst2app}
\end{equation}
where 
\begin{equation}
L_{{\rm max},2}\equiv \frac{n_b N_0 }{\sigma_{\rm tot} \tau_2 n_{\rm IP}} \,
\frac{E_2}{E_1}\, 
\frac{e^
{-\frac{t_{2,{\rm step}}-t_1}{\tau_2}}}{B ( e^{t_1/\tau}-1 )+ 1} \; .
\label{lmax2}
\end{equation}

The integrated luminosity for energy $E_1$ 
stays constant at times $t>t_{2,{\rm step}}$, equal to 
\begin{eqnarray}
\label{se1}
\Sigma_{E_1} & =& \Sigma_1+\Sigma_2 (t_{2,{\rm step}}) +\int_{t_{2,{\rm step}+\Delta t_{\rm ramp}} }^t L_2 (t)  dt  \\ 
& = & 
L_0 \tau \left( \frac{\frac{\Delta Q _{\rm max}}{\Delta Q_0} -1}{1-B } \right) 
+ L_{\rm max} \tau_2 \left( 1 - e^{-\frac{t_{2,{\rm step}} -t_1 }{\tau_2}} \right) \nonumber
 \; ,
\end{eqnarray}
while, for $t>t_{2,{\rm step}}+\Delta t_{\rm ramp}$, the integrated luminosity at energy $E_2$ grows as
\begin{eqnarray}
\label{se2}
\Sigma_{E_2} (t) & =& \int_{t_{2,{\rm step}+\Delta t_{\rm ramp}} }^t L_2 (t)  dt  \\ 
& = & 
 L_{{\rm max},2} \tau_{2,2} 
\left( 1 - e^{
-\frac{t_r-t_{2,{\rm step}} - \Delta t_{\rm ramp}}{\tau_{2,2}} 
} \right) \nonumber
 \; . 
\end{eqnarray}
To determine the ideal run time $t_{r,2}$, now we can either optimize the total integrated luminosity $(\Sigma_{E_1} + \Sigma_{E_2})$ or the luminosity at  $E_2$, namely $\Sigma_{E_2}$.
Let us consider the second approach.

The average luminosity at energy $E_2$ depends on the run time $t_{r,2}$ 
as
\begin{equation}
L_{{\rm ave},E_2} (t_{r,2} )=\frac{\Sigma_{E_2} (t_{r,2} )}{t_{r,2}+t_a }\; .
\label{lavetr2c}
\end{equation}

In analogy to Eqs.~(\ref{Ddef}),
we define auxiliary parameters
\begin{eqnarray}
C_2 & \equiv & 
 L_{{\rm max},2} \tau_{2,2} 
\nonumber
\\
D_2 & \equiv & 
- L_{{\rm max},2} \tau_{2,2} 
e^{\frac{t_{2,{\rm step}} + \Delta t_{\rm ramp}}{\tau_{2,2}}}  \; ,   
\label{D2def2}
\end{eqnarray}
and compute the 
optimum run time 
$t_{r,2}$  
 as 
\begin{equation}
D_2 \left( t_{r,2,{\rm opt}}+t_a +\tau_{2,2} \right) 
e^{-t_{r,2,{\rm opt}}/\tau_{2,2} } + \tau_{2,2} C_2 \stackrel{!}{=} 0 \; ,
\end{equation}
which can again be solved numerically. 

\section{Continuous leveling below the beam-beam limit}
\label{app:ueq} 
In Section \ref{sec:cont}, we considered continuous leveling under the simplifying assumption of a constant beam-beam tune shift.  
However, at the start of the physics fill the total beam-beam 
tune shift may not yet have reached the beam-beam limit,
which could be exploited to further boost the integrated
luminosity. To do so, the emittance 
$\varepsilon$ could initially 
be allowed to shrink as in Eq.~(\ref{emit1}).  
A difference from the fixed-field case 
is that now the damping time 
changes with beam energy as 
\begin{equation}
\frac{1}{\tau (t)} =
\frac{1}{\tau_0} \frac{E^3}{E_{0}^3} 
\; .
\label{tau1c} 
\end{equation} 
In this regime, $dN_b/dt$ is given by Eq.~(\ref{dnbdt2}), 
resulting in 
\begin{equation}
\frac{dE}{dt} = 
 \frac{K}{4 \varepsilon_0}\; E(t)\; N_b(t) \; e^{t/\tau (t)} \; .
\label{dedt2} 
\end{equation}
Using (\ref{enbrel}) and introducing 
$u\equiv E(t)/E_0$,
 this can be rewritten as 
\begin{equation}
\frac{du}{dt} = 
 \frac{KN_0}{4 \varepsilon_0}\; \frac{1}{u^3(t)}\; e^{t u^3(t) /\tau_0}  \; .
\label{dudt} 
\end{equation}
Equation (\ref{dudt}) can be solved numerically.

\end{document}